\documentclass[journal=jpclcd,manuscript=article]{achemso}

\usepackage[version=3]{mhchem}

\usepackage[T1]{fontenc}
\usepackage[english]{babel}

\usepackage{natbib}
\usepackage{mciteplus}
\usepackage{color}
\setcitestyle{numbers,round}

\usepackage[latin1]{inputenc} 
\usepackage{mathtools}
\usepackage{amsmath}
\usepackage{amsfonts}
\usepackage{amssymb} 
\usepackage{amsthm} 
\usepackage{graphicx} 
\usepackage{verbatim} 
\usepackage{enumerate} 
\usepackage{setspace} 
\usepackage{booktabs} 

\newcommand{\be}{\begin{equation}}
\newcommand{\ee}{\end{equation}}
\newcommand{\bea}{\begin{equation*}}
\newcommand{\eea}{\end{equation*}}
\newcommand{\ba}{\begin{array}}
	\newcommand{\ea}{\end{array}}
\newcommand{\beqa}{\begin{eqnarray}}
\newcommand{\eeqa}{\end{eqnarray}}
\newcommand{\beqaa}{\begin{eqnarray*}}
	\newcommand{\eeqaa}{\end{eqnarray*}}

\newcommand{\matr}{\left( \begin{array}}
	\newcommand{\ematr}{\end{array} \right)}

\newcommand{\rb}{\mbox{\boldmath $r$}}

\newcommand{\av}{\mbox{$\vec{a}$}}

\newcommand{\lsim}{{\;\raise0.3ex\hbox{$<$\kern-0.75em\raise-1.1ex\hbox{$\sim$}}
		\;}}
\newcommand{\gsim}{{\;\raise0.3ex\hbox{$>$\kern-0.75em\raise-1.1ex\hbox{$\sim$}}
		\;}}

\def\fwhm{\textsc{fwhm}}
\def\zh{\hat{z}}
\def\rh{\hat{r}}
\def\eadhesion{\varepsilon_\text{adh}}
\def\bear{\begin{eqnarray}}
\def\eear{\end{eqnarray}}

\def\eVnm{eV/nm$^2$}

\newcommand{\citeu}[1]{\textsuperscript{[x]}}

\author{Pekka Koskinen}
\affiliation{Nanoscience Center, Department of Physics, University of Jyv\"askyl\"a, 40014 Jyv\"askyl\"a, Finland}
\email{pekka.j.koskinen@jyu.fi}

\author{Karoliina Karppinen}
\affiliation{Nanoscience Center, Department of Chemistry, University of Jyv\"askyl\"a, 40014 Jyv\"askyl\"a, Finland}

\author{Pasi Myllyperki\"o}
\affiliation{Nanoscience Center, Department of Chemistry, University of Jyv\"askyl\"a, 40014 Jyv\"askyl\"a, Finland}

\author{Vesa-Matti Hiltunen}
\affiliation{Nanoscience Center, Department of Physics, University of Jyv\"askyl\"a, 40014 Jyv\"askyl\"a, Finland}

\author{Andreas Johansson}
\affiliation{Nanoscience Center, Department of Chemistry, University of Jyv\"askyl\"a, 40014 Jyv\"askyl\"a, Finland}
\alsoaffiliation{Nanoscience Center, Department of Physics, University of Jyv\"askyl\"a, 40014 Jyv\"askyl\"a, Finland}

\author{Mika Pettersson}
\affiliation{Nanoscience Center, Department of Chemistry, University of Jyv\"askyl\"a, 40014 Jyv\"askyl\"a, Finland}

\title{Optically Forged Diffraction-Unlimited Ripples in Graphene}

\begin{document}
\begin{tocentry}
\includegraphics[width=2in]{TOC.pdf}
\end{tocentry}

\begin{abstract}
In nanofabrication, just as in any other craft, the scale of spatial details is limited by the dimensions of the tool at hand. For example, the smallest details for direct laser writing with far-field light are set by the diffraction limit, which is approximately half of the used wavelength. In this work, we overcome this universal assertion by optically forging graphene ripples that show features with dimensions unlimited by diffraction. Thin sheet elasticity simulations suggest that the scaled-down ripples originate from the interplay between substrate adhesion, in-plane strain, and circular symmetry. The optical forging technique thus offers an accurate way to modify and shape two-dimensional materials and facilitates the creation of controllable nanostructures for plasmonics, resonators, and nano-optics.
\end{abstract}

\maketitle

One of the central aims in nanoscience is to be able to modify nanostructures at will. Modifications are necessary because it is rarely the pristine materials but the modified and engineered materials that establish functionalities for practical applications.\cite{Stankovich2006,Duong2017} Modifications are particularly necessary for two-dimensional (2D) materials.\cite{Geim2007,Geim2013} Graphene, for instance, gains specific functionalities once modified into ribbons,\cite{Son2006,koskinen_APL_11} introduced with pores or adsorbants,\cite{GomesdaRocha2013,Zhao2014,Koskinen2015,Antikainen2017} or curved into three-dimensional shapes.\cite{castro_neto_RMP_09,Levy2010,Korhonen2014a}

However, all modification techniques have their limitations. Direct mechanical manipulation is either slow and accurate,\cite{Eigler1990} or fast, coarse, and non-reproducible.\cite{kawai_PRB_09,kit_PRB_12} Thermal annealing,\cite{chuvilin_nchem_10,Wang2016} electron irradiation,\cite{kotakoski_PRL_11,Kotakoski2011a} chemical treatment,\cite{thompson-flagg_EPL_09,Svatek2014} and Joule heating\cite{jia_science_09} may be scalable but spatially imprecise due to their random character. It is particularly challenging to modify 2D materials into customized ripples and other 3D shapes. Such modifications frequently require dedicated experimental apparatuses\cite{Bao2009} or specially prepared substrates.\cite{Yamamoto2012a} The difficulty for 3D modification lies partly in substrate adhesion. Although often of weak van der Waals type, adhesion effectively prevents controlled detachment of 2D membranes from the substrate.

Limitations exist also in optical patterning. Although optical techniques may be scalable and easy to apply, the spatial details are determined by the size of the focused laser beam. Creating patterns with details finer than beam size is just as difficult as scribbling equations on a piece of paper with a spray can. Still, optical techniques have plenty of potential to explore, since irradiation provides various mechanisms to modify 2D materials, depending on laser energy and ambient atmosphere.\cite{Aumanen2015,Koivistoinen2016} One particularly promising, still mostly untapped technique is the so-called optical forging, which alone enables controlled and on the fly 3D shaping of graphene.\cite{Johansson2017}

Given the ubiquity of various limitations, there is urgency to improve techniques to modify and engineer 2D materials scalably, accurately, and preferably \emph{in situ}, without customized preparations.

In this work, we demonstrate optical forging of graphene into circular ripples with features much smaller than the size of the laser beam. By using thin sheet elasticity simulations, the rippling is shown to arise from the interplay between substrate adhesion, in-plane stress due to optical forging, and the underlying circular symmetry. Being based on direct irradiation of graphene without specially prepared experimental settings, optical forging provides a practical technique and thereby broadens substantially our abilities to modify and enhance the functionalities of graphene and maybe even other 2D materials. 

To prepare the sample, we grew single-layer graphene by chemical vapor deposition (CVD) on a Cu substrate \cite{Miller2013} and  transferred it to thermally grown SiO$_2$. For fabrication details and graphene characterization, see Supporting Information (SI).

Selected points in the sample were then irradiated by a $515$-nm femtosecond laser focused with an objective lens (N.A. of $0.8$) to a single Gaussian spot. To prevent photoinduced oxidation during the irradiation, the sample was installed inside a closed chamber purged with N$_2$.\cite{Koivistoinen2016} The laser produced $250$~fs pulses at $5-25$ pJ/pulse energy and $600$~kHz repetition rate for a tunable irradiation time $\tau$. This process is called optical forging and results in blistering of the graphene membrane (Fig.~1a and Movie 1 in Supporting Information). Blistering occurs due to local expansion of the membrane, caused by laser-induced defects and the related compressive in-plane stress.\cite{Johansson2017} The local expansion field $\varepsilon(\rb)$ therefore depends on the time-integrated laser intensity profile $I(\rb)$, which enables accurate control over the expansion and blister height via the irradiation time $\tau$. Consequently, we irradiated the sample at separate spots for irradiation times ranging from $\tau=0.1$ to $3600$~seconds. Finally, the blistered sample was characterized by Raman spectroscopy and measured by an atomic force microscope (AFM; see SI).

\begin{figure}[tb]
	\includegraphics[width=\columnwidth]{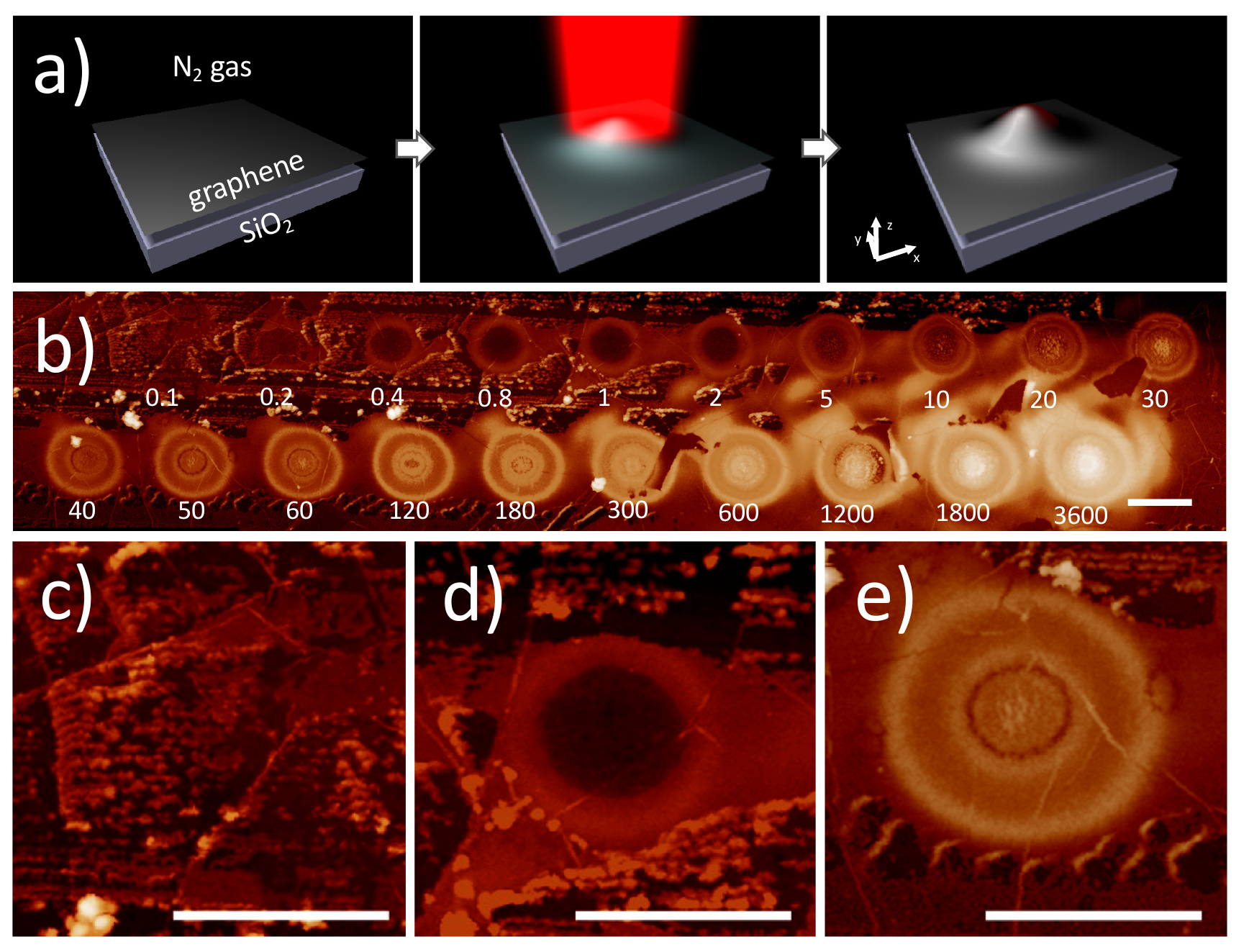}
	\caption{Monitoring the gradual formation of optically forged graphene blisters on SiO$_2$. a) In optical forging, graphene is irradiated by focused femtosecond laser beam under inert N$_2$ atmosphere. The laser creates defects that cause isotropic expansion of graphene membrane and trigger the formation of blisters. The blisters are hollow and not pressurized.\cite{Johansson2017} b) Atomic force microscope image of blisters formed at progressively increasing irradiation time $\tau$ (numbers show $\tau$ in seconds; highest features are $60$ nm). Blisters form at $\tau>0.4$~s, initially with one circular ripple, later with several ripples and a dome in the center. c) Zoom into an irradiated area with $\tau=0.2$~s, where the graphene still remains flat. Visible are only the patchy residues from sample processing. d) Zoom into a blister with one ripple ($\tau=1$~s). e) Zoom into a blister with multiple ripples ($\tau=50$~s). Scale bars, $1$~$\mu$m.}
	\label{fig:blisters}
\end{figure}

The systematic increase in irradiation time produced a nontrivial but beautiful and reproducible pattern of blisters (Figs~1b and S4). In particular, blisters had profiles more complex than the usual domes.\cite{koenig_nnano_11} At short irradiation times ($\tau<0.4$~s) the graphene remained flat on the substrate (Fig.~1c). At intermediate irradiation times ($0.4\leq \tau \leq 2$~s) the graphene developed blisters with one circular ripple (Fig.~1d). At long irradiation times ($\tau \geq 5$~s) the graphene developed concentric ripples in progressively increasing numbers and a gradually developing central dome (Fig.~1e and Movie 2 in Supporting Information). Parts of the area between the blisters were detached from the substrate because the laser irradiated also during the movement from one spot to another. Note that the radial features in the ripples have dimensions down to $100$~nm, nearly \emph{ten times smaller} than the laser spot and the ripple diameters themselves. Optical forging can thus reach 3D shaping of graphene that beats the diffraction limit. This is our main result.


To quantify the expansion of the graphene membrane, we used AFM height profiles to measure the increase in the surface area of the blisters. Within the projected areas of about $1$~$\mu$m$^2$, the corrugated membrane areas increase nearly monotonously upon increasing irradiation time, reaching $10^{-2}$~$\mu$m$^2$ ($\sim 1$\%) area increase at $\tau=1$~hour (Fig.~2a). Initially, the area increases linearly in irradiation time, at rate $22$~nm$^2$/s (Fig.~2b). This area increase was used to determine radius-dependent linear expansion, $\varepsilon(r)$. By assuming here a one-photon process and a Gaussian laser intensity profile $I(r)$, we obtain 
\begin{equation}
\varepsilon(r)=\varepsilon_0 \exp{\left(-4r^2\log{2}/\fwhm^2 \right)},
\label{eq:epsr}
\end{equation}
where $r=0$ at the center of the spot and $\fwhm=800$~nm is the full width at half maximum of the laser beam. Because the laser focal spot was difficult to maintain, $\fwhm$ had to be treated as a parameter and adjusted to give the best overall fit to the observed lateral dimensions in the experiment. The maximal expansion $\varepsilon_0$ increases at the rate $1.5\times 10^{-3}$~\%/s at short irradiation times and saturates at almost $1$~\% at long irradiation times (Fig.~2). The initial linear rate and the saturation are in good agreement with earlier experiments.\cite{Johansson2017}

\begin{figure}[t!]
	\includegraphics[width=\columnwidth]{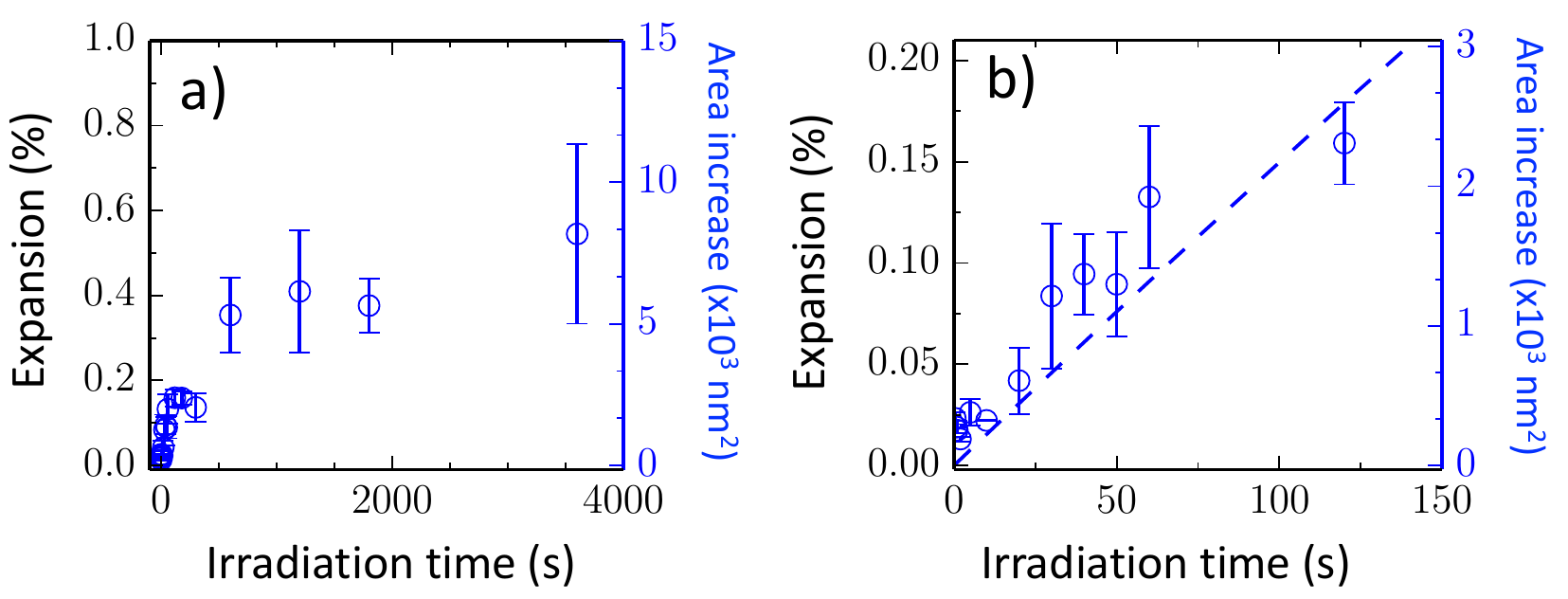}
	\caption{Effective expansion of graphene membrane during laser irradiation. a) Area increase due to blister formation, as measured from the blister profiles of Fig.~1b (right scale). Area increase transformed into maximum linear expansion in the middle of the laser spot (left scale). b) Zoom into $\tau<150$~s. A linear fit gives an expansion rate $1.5\times 10^{-3}$~\%/s or $22$~nm$^2$/s (dashed line). The vertical bars are uncertainties in blister areas.\cite{errors}}
	\label{fig:strains}
\end{figure}

The diffraction-unlimited rippling suggests a mechanism that involves competition between surface adhesion and expansion-induced stress. To investigate the mechanism in detail, we simulate blister growth by classical thin sheet elasticity model.\cite{landau_lifshitz} Such models have proven successful in the modeling of deformed graphene membranes. \cite{kudin_PRB_01,Shenoy2008,Shenoy2010,kit_PRB_12,Lambin2014,Koskinen2016} The energy in the model contains in-plane strain energy, out-of-plane bending energy, and surface adhesion. The laser-induced isotropic expansion is introduced via the diagonal of the in-plane strain tensor as $e_{\alpha \beta}(\rb)=e^0_{\alpha \beta}(\rb) - \delta_{\alpha \beta} \varepsilon(\rb)$, where $\varepsilon(\rb)$ is the expansion field and $e^0_{\alpha \beta}(\rb)$ is the strain tensor of the unexpanded, pristine graphene.\cite{Johansson2017} The adhesion is modeled by the generic $12-6$ Lennard-Jones potential.\cite{Koskinen2014} This model was discretized, implemented in two computer codes (with and without circular symmetry), and used to optimize blister geometries for given $\varepsilon_0$ and adhesion energy $\eadhesion$.\cite{bitzek_PRL_06} For details, see Supporting Information.

Before analyzing the model in full, it is instructive first to ignore adhesion and calculate few analytical results. Due to the smallness of the graphene bending modulus, at $\mu$m-length scales the mechanical behavior is dominated by in-plane strain energy.\cite{Korhonen2015} The strain energy is minimized when $e_{\alpha \beta}\approx 0$, or $e^0_{\alpha \beta}\approx \delta_{\alpha \beta}\varepsilon(\rb)$. To a first approximation, Eq.~(\ref{eq:epsr}) then implies an area increase of $\Delta A = [\pi/(2\log{2})] \times \fwhm^2  \varepsilon_0$. (This relation was used earlier to transform $\Delta A$ into $\varepsilon_0$.) With the displacement vector $\av(r)=a_r(r) \rh + a_z(r) \zh$, the diagonal components of the strain tensor become
\begin{align}
\begin{cases}
e_{rr}(r)= a'_r(r) + \frac{1}{2}\left[{a'_r(r)}^2+{a'_z(r)}^2\right]-\varepsilon(r)\\
e_{tt}(r)= a_r(r)/r + \frac{1}{2}\left[a_r(r)/r\right]^2-\varepsilon(r), \\
\end{cases}
\end{align}
where $r$ refers to radial and $t$ to tangential in-plane component, and prime stands for a derivative with respect to $r$. Since the in-plane strain energy minimizes at $e_{\alpha \beta}\approx 0$, we obtain $a_r(r)\approx r \varepsilon(r)$ and 
\begin{equation}
a'_z(r)\approx \pm \sqrt{16\log 2 \times \varepsilon(r)}(r/\fwhm).
\label{eq:slope}
\end{equation}
That is, when the membrane adapts to isotrotopic expansion under radial symmetry, energy gets minimized by adjusting the \emph{slope into a fixed absolue value}. When the slope is negative for all $r$, integration yields the profile $a_z(r)=\fwhm\times \sqrt{\varepsilon(r)/\log{2}}$. This profile corresponds to a blister with one central dome and a maximum height of $h_{max}=\fwhm\times\sqrt{\varepsilon_0/\log{2}}$. Numerically optimized blister profile follows this analytical estimate accurately (Fig.~3a).

\begin{figure}[t]
	\includegraphics[width=\columnwidth]{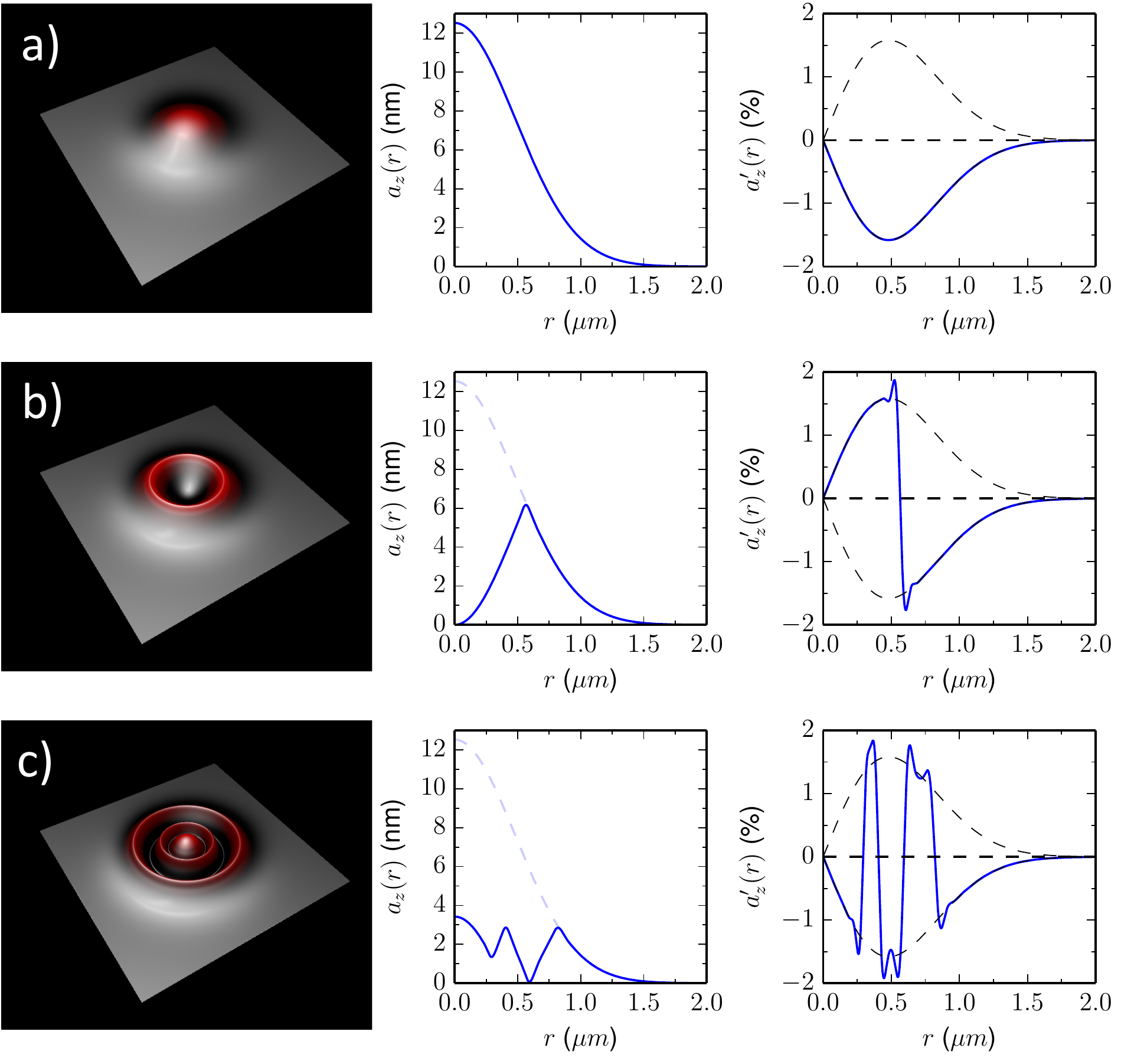}
	\caption{Thin sheet elasticity modeling of blisters with $\varepsilon_0=0.017$~\% and $\eadhesion=0$ (no substrate adhesion). a) Blister with one central dome. b) Blister with one circular ripple. c) Blister with two concentric ripples and a central dome. Panels show visualizations (left; height exaggerated), radial height profiles $a_z(r)$ (middle), and the slopes of the radial height profiles $a_z'(r)$ (right). Dashed lines on the right show the analytical limits for $a_z'(r)$ from Eq.(\ref{eq:slope}).}
	\label{fig:noadhesion}
\end{figure}

However, positive and negative slopes in Eq.~(\ref{eq:slope}) are equally acceptable. Since the energy cost of bending is small, it is cheap to create a kink that reverses the sign of $a_z'(r)$ abruptly. This kink appears topologically as a perfectly round ripple (Fig.~3b). Multiple kinks at different radii produce concentric ripples of varying heights and diameters (Fig.~3c). Compared with the scale of in-plane strain energy, blisters of different ripple counts are nearly isoenergetic. When the number of ripples increases, the slopes progressively deviate from Eq.~(\ref{eq:slope}). Otherwise, the analytical description of the blister profiles without adhesion is apparent.




The role of adhesion, then, is to pull the membrane down, toward the substrate. Understanding the behavior of adhesion-free membranes is helpful, but when elastic and adhesive energies compete, we have to rely on numerical simulations. We took a closer look at the blister with $\tau=1$~s, which is near the onset of blistering (Fig.~1d). This $4$-nm-high blister has a $0.97$~$\mu$m ripple diameter and $\varepsilon_0=0.017$~\%\ expansion, as given by the AFM profile. We simulated this blister using the experimental $\varepsilon_0$ and adhesion in the range $\eadhesion=0\ldots 1$~\eVnm. 

When $\eadhesion<1$~$\mu$eV/nm$^2$, the ripple is broad and the middle of the blister is mostly detached from the substrate, disagreeing with the experiment (Fig.~4a); adhesion remains a minor perturbation to the zero-adhesion profile (Fig.~3b). When $\eadhesion>10$~$\mu$eV/nm$^2$, in turn, the ripple becomes too narrow and shallow, also disagreeing with the experiment; when $\eadhesion \gtrsim 100$~$\mu$\eVnm, the membrane ultimately snaps flat on the substrate. However, around $\eadhesion\approx 3$~$\mu$\eVnm, adhesion pulls the membrane down so that both the ripple width and height agree with the experiment. Using the adhesion $\eadhesion=3$~$\mu$\eVnm, one-ripple blistering occurs at $\varepsilon_0\approx 0.005$~\%\, and the blister height increases linearly when $\varepsilon_0$ increases further (Fig.~4b). This simulated trend agrees with the experimental trend in one-ripple blisters ($\tau \lesssim 2$~s). These agreements suggest that the adhesion between laser-modified graphene and SiO$_2$ is observable but substantially smaller than typically observed for pristine van der Waals solids and clean interfaces.\cite{bjorkman_PRL_12}

For completeness, we optimized all $18$ blisters by using $\eadhesion=3$~$\mu$\eVnm\ and by adopting the observed set of ripples as initial guesses. After optimization, the resulting pattern of blisters turned out similar to the experimental ones (Fig.~4f). At small $\varepsilon_0$ stable blisters have only one ripple (Fig.~4c), but at larger $\varepsilon_0$ stable blisters have multiple ripples (Fig.~4d). Simulations capture the main features of the experimental blisters, even if they deviate with respect to certain details, presumably due to asymmetric expansion field and small variations in initial conditions of the graphene membrane, generated during the sample fabrication.



\begin{figure}
	\includegraphics[width=1\columnwidth]{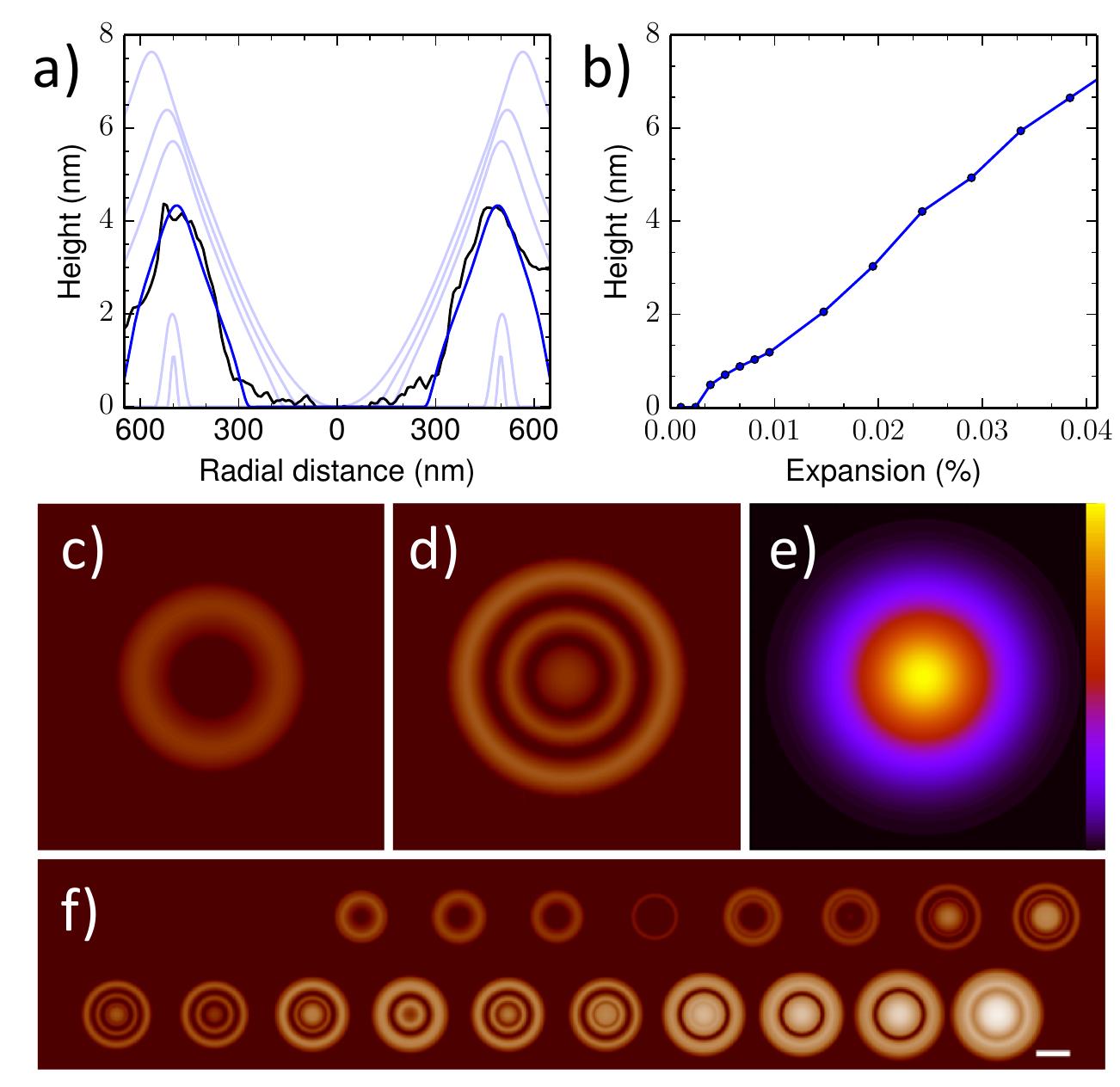}
	\caption{Thin sheet elasticity modeling of blisters with adhesion. a) The experimental profile of $\tau=1$~s blister (black curve) compared to simulated profiles of one-ripple blisters with different adhesions (blue curves from top to bottom: $\eadhesion=0$, $0.5$, $1.0$, $3.0$, $10$, and $100$ $\mu$eV/nm$^2$). b) The height of one-ripple blister as a function of expansion $\varepsilon_0$. c) Contour plot of a one-ripple blister with $\varepsilon_0=0.017$~\% (corresponding to $\tau=1$~s, Fig.~1d). d) Contour plot of a multiple-ripple blister with $\varepsilon_0=0.09$~\% (corresponding to $\tau=50$~s, Fig.~1e). e) Contour plot of $\varepsilon(r)/\varepsilon_0$ for all blisters. The color scale is linear from zero to one. f) Contour plots for all energy-optimized blisters, using the expansions from Fig.~2b and the initial guesses from Fig.~1b. Scale bar, $1$~$\mu$m. Field of view in panels c-e is $2.2\times 2.2$~$\mu$m$^2$. Panels b-f have $\eadhesion=3$~$\mu$eV/nm$^2$ and all blisters are optimized without imposing radial symmetry.}
	\label{fig:quantify}
\end{figure}


Yet a question remains: why do blisters initially appear with one circular ripple? This question can be addressed by considering Eq.~(\ref{eq:slope}). The preferred slope has a maximum at $r_0=\fwhm/2\sqrt{\ln{2}}=0.48$~$\mu$m. In other words, around radius $r_0$ the energy to keep the membrane flat is the largest. When the in-plane stress in a flat membrane increases upon increasing $\varepsilon_0$, it becomes energetically favorable to release the stress by \emph{creating the kink right at $r_0$} and making a circular ripple with diameter $2r_0=0.96$~$\mu$m. This result agrees well with the observations. Upon continuous irradiation, after the initial ripple has appeared, the ripple height increases until it becomes energetically favorable to create more ripples. This implies a process-dependent rippling of ever-increasing complexity.

This scenario for rippling was confirmed by performing global optimizations for blisters with $\eadhesion=0.1-100$~$\mu$\eVnm, $\varepsilon_0=0.001\ldots 1$~\%, and various types of initial guesses. First, at sufficiently small $\varepsilon_0$, the membrane remains flat without blistering. A critical limit for blistering is around $\varepsilon_0^c \approx 0.02\times (\eadhesion/\text{ eVnm}^{-2})^{1/2}$. Second, when $\varepsilon_0$ increases just above the critical limit, the first blisters \emph{always} have one ripple with diameter $D_0\sim 1$~$\mu$m, independent of $\eadhesion$. This result is in agreement with the experiments and with the maximum-slope argument given above ($D_0\approx 2r_0$). Third, at intermediate values of $\varepsilon_0$, blisters show a complex pattern of ripples of varying heights and diameters. Fourth, at the limit of large $\varepsilon_0$, the in-plane strain energy dominates and the minimum energy blisters always have one central dome (Fig.~3a).

Compared to the typical magnitude of adhesion ($1-2$~eV/nm$^2$) between clean interfaces of van der Waals solids, \cite{vanin_PRB_10,Hamada2010,bjorkman_PRL_12,Tang2014a} the adhesion in the model ($\sim 1$~$\mu$\eVnm) is small. The smallness, however, is apparent even in a back-of-the-envelope calculation. Namely, upon blistering, the gain in elastic energy density is $k_s\varepsilon_0^2/(1-\nu)$ and the cost of adhesion energy density is $\eadhesion$. At the onset of blistering, the two energies are equal, $\eadhesion\sim k_s\varepsilon_0^2/(1-\nu)$. Since the blisters appear at $\varepsilon_0\sim 0.02$~\%, the adhesion has to be around $1-10$~$\mu$\eVnm. The small adhesion may be due to water or functional groups \cite{Gao2014}, topographic corrections,\cite{Delrio2005} electrostatics due to localized charge traps,\cite{miwa_APL_11} or other experimental details.\cite{Boddeti2013,Miskin2017} A detailed investigation of the laser-modified adhesion will be pursued later.

To summarize, by using the optical forging technique, we created diffraction-unlimited circular ripples in graphene on SiO$_2$. The rippling could be explained by the presence of circular symmetry amid the competition between substrate adhesion and in-plane compressive stress. In other words, the tiny rippling results \emph{spontaneously} after creating an inhomogeneous expansion field at much larger length scale. We can therefore straightforwardly predict that upon shrinking the size of the laser beam, the ripples will get smaller still. Once the mechanism responsible for the expansion of graphene is understood better, the technique could be applied also to other substrates and 2D materials. However, already now the technique and our observations provide many openings for novel research. A straightforward extension will be to control the rippling by engineering beam shapes. The technique produces beautiful circular blisters that probably have well-defined vibrational frequencies and can be used in resonators.\cite{Bunch2007} Via formation of circular ripples, the technique also produced controllable curvatures that can be used to launch localized plasmons.\cite{Smirnova2016} Thus, in addition to producing new physics and posing fundamental questions such as that of the laser-modified adhesion, the technique opens new avenues in the research of two-dimensional materials.

{\bf Acknowledgements.}
We acknowledge the Academy of Finland for funding (projects 297115 and 311330).

{\bf Supporting Information available.}
Supporting Information presents details of sample characterization and computer simulations and two movies illustrating the blister formation.

The authors declare no competing financial interest.


\begin{mcitethebibliography}{52}
\bibitem[err()]{errors}
	Uncertainties are obtained by calculating areas in two ways: i) assuming full
	circular symmetry by using one radial height profile starting from the center
	and ii) assuming partial circular symmetry by using one height profile across
	the entire blister.\relax
	\mciteBstWouldAddEndPunctfalse
	\mciteSetBstMidEndSepPunct{\mcitedefaultmidpunct}
	{}{\mcitedefaultseppunct}\relax
	\EndOfBibitem
\end{mcitethebibliography}

\providecommand{\latin}[1]{#1}
\makeatletter
\providecommand{\doi}
{\begingroup\let\do\@makeother\dospecials
	\catcode`\{=1 \catcode`\}=2\doi@aux}
\providecommand{\doi@aux}[1]{\endgroup\texttt{#1}}
\makeatother
\providecommand*\mcitethebibliography{\thebibliography}
\csname @ifundefined\endcsname{endmcitethebibliography}
{\let\endmcitethebibliography\endthebibliography}{}

\end{document}